\documentstyle[11pt,newpasp,twoside,epsf]{article}
\def\edcomment#1{\iffalse\marginpar{\raggedright\sl#1\/}\else\relax\fi}
\reversemarginpar

\begin{document}
\title{Structure and Conditions in Massive Star Forming Giant Molecular Clouds}
\author{Jonathan Williams}
\affil{Astronomy Department, University of Florida, 211 Bryant Space Science Center, Gainesville FL, USA; williams@astro.ufl.edu}

\begin{abstract}
Massive stars form in clusters within self-gravitating molecular clouds.
The size scale of these clusters is sufficiently large that non-thermal,
or turbulent, motions of the gas must be taken into account when
considering their formation. Millimeter wavelength radio observations
of the gas and dust in these clouds reveal a complex, self-similar
structure that reflects the turbulent nature of the gas. Differences
are seen, however, towards dense bound cores in proto-clusters.
Examination of the kinematics of gas around such
cores suggests that dissipation of turbulence may be the first
step in the star formation process. Newly formed stars, on the
other hand, replenish turbulence through their winds and outflows.
In this way, star formation may be self-regulated.
Observations and simulations are beginning to demonstrate the key
role that cloud turbulence plays in the formation and evolution of
stellar groups.
\end{abstract}

\section{Molecular cloud structure}
Massive stars form in Giant Molecular Clouds (GMCs) that have
masses $M>10^5~M_\odot$, sizes $R\simeq 20-50$~pc,
linewidths $\Delta v_{\rm FWHM}\sim 2-5$~km~s$^{-1}$
and kinetic temperatures $T\simeq 10-20$~K.
Since the observed linewidths are a factor $5-10$ greater
than the thermal linewidth for such cool objects, the gas
motions within GMCs are highly turbulent.
Such turbulent conditions manifest themselves not only through the
high spectral linewidths but also in the complex internal structure
of these clouds (Williams, Blitz, \& McKee 2000).

Molecular cloud structures have been studied over a wide range of
scales and environments 
and scaling laws such as the size-linewidth relation and mass spectrum
are found to be amazingly similar from cloud to cloud
{\it independent of their star forming nature}. Such a high
degree of self-similarity lends itself to fractal models of clouds
but cannot be directly related to star formation since the same
structures are observed in clouds with stars and those without.

Recent observations of thermal emission from dense, dusty condensations
in star forming regions, however, have shown departures from self-similarity.
Motte, Andr\'{e}, \& Neri (1998) and Testi \& Sargent (1998)
found that the mass spectrum, $dN/dM\sim M^{-\alpha}$,
of cores within cluster forming regions was steeper ($\alpha > 2$)
than typically measured for structures within clouds ($\alpha = 1.5-1.8$)
and closer to the stellar IMF ($\alpha = 2.35$).

The difference between these dust continuum observations and the
molecular spectral line observations are that the former are of
dense, bound, individual star forming cores while the latter
are of lower (average) density structures. Moreover, the
linewidths of the dust continuum cores are small, near the
thermal value, while the spectral line clumps are predominantly
non-thermal. If we equate the structural similarities
of the clumps with the universal nature of cloud turbulence
then is the departure from self-similarity in the thermally supported
cores related to the loss of this turbulent support?
Examining turbulence in molecular clouds may lead to a physical
understanding of the relation between cloud structure and star formation.

\section{The role of turbulence in star formation}

Most stars, and particularly all massive stars, form in clusters
over size scales, $>0.2$~pc, where the pre-star forming material
is supported by turbulent motions of the gas.
However, numerical simulations of hydrodynamic and magnetohydrodynamic
turbulence in GMCs show that such motions cannot be maintained
over more than a few free-fall times (Mac Low 1999),
yet the observed low star formation efficiency of
molecular clouds requires that cloud support be quasi-static.
We are led, therefore, to a dynamic picture of molecular clouds
in which turbulent motions are in a continual state of dissipation
and replenishment. Observations of the velocity field in 
the Serpens cloud show both these effects.

The NE region of the Serpens molecular cloud contains a
deeply embedded cluster of very young (Class 0) protostars.
It is one of the nearest known examples of cluster formation
and although it only contains low mass stars, it provides an
opportunity to study how stars form in groups and is
therefore an important stepping stone for understanding
massive star formation.
Here I briefly summarize BIMA $\lambda$3~mm
inteferometer observations of the dense gas toward the cluster.
Full details can be found in Williams \& Myers (2000).

Observations were made in two molecular lines; the optically thin
N$_2$H$^+$(1--0) tracing the turbulent velocity field of the gas,
and the optically thick CS(2--1) tracing outer core envelopes and
used as a diagnostic of infall and outflow motions. Several ``quiescent
cores'' were found in a map of non-thermal N$_2$H$^+$ velocity dispersion.
These represent localized regions of turbulent dissipation.
Conversely, the non-thermal N$_2$H$^+$ velocity dispersion reached a
maximum around the two brightest embedded protostars indicating
local stirring of the velocity field (this applies to the dense
gas in the cores since N$_2$H$^+$ is not seen in protostellar outflow wings).
The CS line was generally self-absorbed across the cluster and, by
modeling the asymmetry in the double-peaked profile, it was possible
to determine the relative inward/outward motions of the cores.
Inward motions were greatest toward the quiescent cores and reversed
around the regions of greatest N$_2$H$^+$ linewidth. This correlation
suggests that the inward and outward motions are turbulent flows
from high to low pressure (linewidth).

The initial growth and contraction of star forming cores may occur,
therefore, through the loss of turbulent pressure support.
This process is more dynamic than the quasi-static slippage of
neutral particles through magnetic field lines (although
ambipolar diffusion may still characterize the last stages of
core collapse).
Newly formed stars stir up the gas through their powerful winds
and outflows and can reverse the pressure gradient and subsequent
flow. Perhaps star formation becomes self-regulated in this manner
(c.f. Norman \& Silk 1980).

New observations of a second, less embedded and hence more evolved,
cluster in the SE region of the Serpens cloud show evolutionary
differences in comparison with the NW cluster.
Several quiescent cores were again seen in the N$_2$H$^+$ line
but the core-core velocity dispersion, measured over a similar
area, was higher in the more evolved cluster (Figure~1)
and similar to that measured for small stellar groups (Jones 1971).
The evolution from a tightly bound group to a more loosely
bound, and ultimately unbound, state is expected as a protocluster
emerges from its natal molecular cloud (Hills 1980).
Detailed observations of individual star forming cores within
embedded clusters, such as described here, should reveal the
processes involved.

\begin{figure}
\plotfiddle{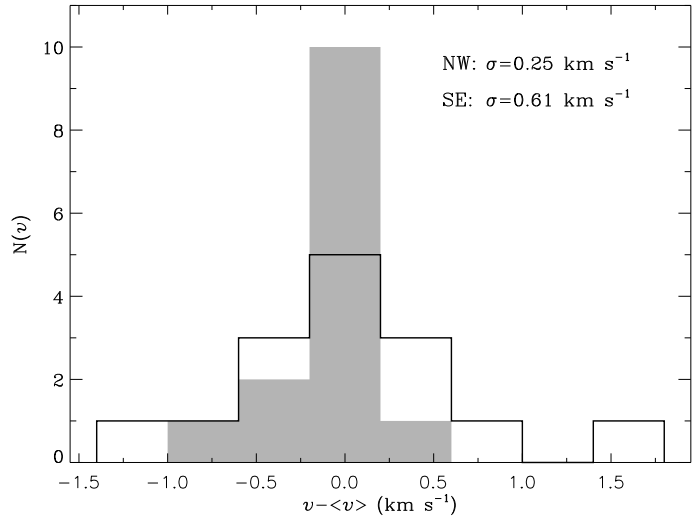}{2.45in}{90}{38}{38}{160}{-19}
\caption{Core-core velocity dispersion in two protoclusters in the
Serpens molecular cloud. The shaded histogram represents the distribution
of core velocities about the mean for a deeply embedded group of
Class 0 sources in the NW region of the cloud, and the solid line
shows the distribution for a less embedded cluster in the SE.
As a cluster emerges from a molecular cloud the loss of surrounding
mass causes the stellar group to become less bound.}
\end{figure}

\end{document}